\documentstyle[prl,aps,multicol]{revtex}
\begin{document}

\title{Devil's staircase for a nonconvex interaction}
\author{Janusz J\c{e}drzejewski}

\address{Institute of Theoretical Physics,
University of Wroc{\l}aw
\\
pl. Maksa Borna 9, 50--204 Wroc{\l}aw, Poland,
\thanks{e-mail:
jjed@ift.uni.wroc.pl}
 \\
 and \\
Department of Theoretical Physics, \\
University of {\L}\'{o}d\'{z}\\
ul. Pomorska 149/153, 90--236 {\L}\'{o}d\'{z}, Poland}

\author{Jacek Mi\c{e}kisz}
\address{Institute of Applied Mathematics and Mechanics,\\
University of Warsaw \\
ul. Banacha 2, 02--097 Warszawa, Poland,
\thanks{ e-mail:
miekisz@mimuw.edu.pl}
}
\date{\today}
\maketitle
\begin{abstract} We study ground-state orderings of
particles in classical lattice-gas models of adsorption on crystal
surfaces. In the considered models, the energy of adsorbed particles
is a sum of two components, each one representing the energy of
a one-dimensional lattice gas with two-body interactions
in one of the two orthogonal lattice directions. This feature reduces
the two-dimensional problem to a one-dimensional one.
The interaction energy in each direction is repulsive and strictly
convex only from distance 2 on,
while its value at distance 1 can be positive or negative,
but close to zero.
We show that if the decay rate of the interactions is fast enough, 
then particles form 2-particle lattice-connected aggregates 
which are distributed in the same most homogeneous way
as particles whose interaction is strictly convex everywhere. 
Moreover, despite the lack of convexity,
the density of particles versus the chemical potential appears to be
a fractal curve known as the complete devil's staircase.
\end{abstract}

\begin{multicols}{2}
One of the problems of surface physics is to study orderings 
of atoms (adatoms) adsorbed on crystal surfaces.
The phenomenon can be modelled by certain classical 
lattice gases \cite{ishimura,sasaki,oleksy}.
Adatoms are assumed to sit on top of substrate atoms of the
square lattice \cite{ishimura,sasaki}.
Substrate deformation causes effective, long-range
interactions between adatoms.
In many cases of interest, the Hamiltonian of such system
can be decomposed into two parts; one involving two-body interactions
between particles located along the $x$-direction
and the other one along the perpendicular 
$y$-direction of the square lattice. Such one-dimensional interactions
constitute sometimes certain local perturbations of repulsive 
and strictly convex interactions. Knowing the one-dimensional
ground-state configurations,
we can easily construct two-dimensional ground-state configurations 
of the original model, as observed in \cite{ishimura} 
and discussed in the end of our paper.
Therefore, in what follows we deal exclusively with the one-dimensional
situation.

In classical lattice-gas models, considered in the sequel,
every site of a one-dimensional lattice ${\bf Z}$
can be occupied by one particle or be empty.
Then, a (infinite-lattice) configuration is an assignment of
particles to lattice sites.
We assume that particles interact only through two-body forces
and to a pair of particles at lattice sites $i$ and $j$,
whose distance is $r=|i - j|$, we assign the translation-invariant
{\em interaction energy} $V(r)$. The Hamiltonian of our system
in a bounded region $\Lambda$ can be then written in the following
form:
\begin{equation}
\label{H}
H_{\Lambda}(X)=
\sum_{ \{\{i,j\}: \{i,j\} \cap \Lambda \neq \emptyset \} }
V(|i - j|)s_{i}(X)s_{j}(X),
\end{equation}
where $s_{i}(X)$ assumes value $1$ if in the configuration $X$
the site $i$ is occupied by a particle and otherwise value $0$.

If we assume that $V(r)$ is positive for every $r \geq 1$
and strictly convex, that is $V(r)+V(r+2) > V(r+1)$, then we obtain  
a model used by Hubbard to study orderings of electrons in
quasi-one-dimensional conductors \cite{hub78,hub79} and by Pokrovsky
and Uimin to study orderings of monolayers of atoms
adsorbed on crystal surfaces \cite{pokro}.
For any given particle density $\rho$,
Hubbard found the energy density $e(\rho)$ of ground-state configurations.
He also showed that any ground-state configuration has the
following remarkable property. Let $x_{i} \in {\bf Z}$ be a coordinate
of the $i$-th particle. Then there exists a sequence
of natural numbers $d_{j}$ such that 
$x_{i+j}-x_{i} \in \{d_{j}, d_{j}+1\}$ for every $i \in {\bf Z}$
and $j \in {\bf N}$, i.e., the distances between the first (second, etc.)
nearest neighbors can differ by at most one lattice constant.
Configurations with such property are known as the
{\em generalized Wigner lattices} \cite{hub78} or the
{\em most homogeneous configurations} \cite{lemberger}.

The ground-state phase diagram in the grand-canonical ensemble 
has been calculated heuristically by Bak and Bruinsma \cite{bak}
while a proof for the related Frenkel-Kontorova model
has been provided by Aubry \cite{aub} and adapted to the lattice-gas model
case by Mi\c{e}kisz and Radin \cite{mie1}; it is outlined below.

In the grand-canonical ensemble, with the fixed chemical potential
$\mu$, to find the energy density of a ground state we have to minimize
\begin{equation}
f(\rho)=e(\rho)-\mu \rho.
\end{equation}  
Now, $e(\rho)$ is differentiable at every irrational $\rho$ and is 
nondifferentiable at any rational $\rho$ \cite{aub,mie1}. However,
as a convex function,
it has a left derivative $d^{-}e(\rho)/d\rho$ and a right derivative 
$d^{+}e(\rho)/d\rho$ at every $\rho$. It follows that to have a ground state
with an irrational density $\rho$ of particles, one has to fix
$\mu(\rho)=de(\rho)/d\rho$. For any rational $\rho$, there is 
a closed interval
of chemical potentials $[d^{-}e(\rho)/d\rho, d^{+}e(\rho)/d\rho]$,
where the most homogeneous configurations of density $\rho$
are the ground-state configurations.
One can show that the sum of lengths of these intervals amounts
to the length of the interval begining at the end of the half-line,
where the vacuum is the only ground-state configuration, and
ending at the begining of the hal-line, where the completely filled
configuration is the only ground-state configuration. 
The particle density versus the chemical potential of particles,
$\rho(\mu)$,
is constant on each of the above described intervals. Moreover, 
it is a continuous function on the real line. The curve $\rho(\mu)$
is classified as a fractal one and named the {\em complete devil's staircase}
\cite{bak,aub,mie1}.

For any rational $\rho$, there is a unique
(up to translations) periodic ground-state configuration with that density
of particles. For any irrational
$\rho$, there are uncountably many ground-state configurations which are
the most homogeneous configurations. It has been shown in \cite{mie2} 
that they all belong to one local isomorhism
class. It means that they cannot be locally distinguished one from
another. Every local pattern of particles present in one ground-state
configuration appears in any other within a bounded distance, in other
words (more formally), there exists the unique ground-state measure
supported by them.

However, it appears that in the models of adsorption \cite{ishimura,sasaki},
$V(1)$ is very small (positive, negative or zero) 
as compared with $V(2)$, violating therefore the convexity 
property of interactions. It may seem that the smalness of $V(1)$ together 
with the repelling nature of remaining interactions will force particles
to form 2-particle lattice-connected aggregates, called 2-molecules, 
allowing remaining 
particles to be farther apart. Then, because the interaction between 
2-molecules is strictly convex, we would obtain a molecular devils'staircase.
However, as the folowing example shows, one cannot hope to obtain 
such a universal result, independent of the details of interactions,
as in the case of strictly convex interactions.

\noindent {\bf Example 1}

\noindent Consider the following convex potential $V$ 
which in some units is given by 
$V(1)=0$, $V(2)=4$, $V(3)=2$, $V(4)=1$, and $V(r)=0$ for $r \geq 5$,
and two periodic configurations (period 17) of particle density
$\rho=8/17$ whose elementary cells are of the form:
$[\bullet \bullet \circ \circ \circ \bullet \bullet\bullet
\circ \circ \circ  \bullet \bullet\bullet \circ \circ \circ ]_{1}$
and
$[\bullet \bullet \circ \circ \bullet \bullet \circ \circ
\bullet \bullet \circ \circ \bullet \bullet \circ \circ \circ ]_{2}$,
where $\bullet$ stands for a particle while $\circ$ for an empty
site.
The energy per elementary cell in the first case is
$2 V(2)$ + $3 V(4) = 11$ while in the second case it is
$3 V(3)$ + $7 V(4) = 13$. Thus the configuration $[.]_{2}$ that consists
exclusively of 2-molecules looses against the configuration
$[.]_{1}$ that consists of 2-molecules and 3-molecules.
This remains true if we do not set $V(r)=0$ for $r \geq 5$ but require that 
$V(r)$ be positive and stricly convex from $r=2$ on and sufficiently
fast decaying from $r=5$ on.

Now, in order to get rid of $n$-molecules, $n \geq 3$, we restrict the
family of considered potentials, demanding that $V(2)$ be relatively
strong with respect to all other $V(r)'s$. We have arrived at
the following result:

There are no $n$-molecules with $n \geq 3$ in any ground-state configuration
in a lattice gas (\ref{H}) with an interaction energy $V$
such that $V(1) > - 2V(2)$ and $V(2) \geq (7W+V(1))/2$, 
where $W=\sum_{r=3}^{\infty} V(r)$. 

The idea of the proof is to start with an arbitrary configuration
containing $n$-molecules, removing $n-2$ particles from every
$n$-molecule,
rearranging the resulting configuration and then putting back removed
particles in the form of 2-molecules.
One can show that if the above conditions
are satisfied, then the gain in energy caused by removing
particles is strictly larger than its loss
associated with inserting back the particles.

The following example shows that configurations containing 
isolated particles, called atoms, are unlikely to be the ground-state ones.

\noindent {\bf Example 2}  

\noindent Consider the following periodic (period 13) configuration:
$[\bullet \circ \circ \bullet \bullet \circ \circ \bullet
\circ \circ \bullet \bullet \circ ]_{3}$. The elementary cell $[.]_{3}$
contains a local configuration of two atoms separated by a 2-molecule.
Let us transform this local configuration into another one, consisting
of two 2-molecules only, by moving to each other the left atom and the right
particle of the 2-molecule and the right atom and the right particle of
the 2-molecule. The resulting elementary cell reads:
$[\circ \bullet \bullet \circ \circ \bullet \bullet \circ \circ 
\circ \bullet \bullet \circ ]_{4}$.
One can easily calculate the relative energy change of the local
configurations considered above;
for a potential described in Example 1 it is negative, showing
that the original periodic configuration is not a ground-state one.

We have generalized the above procedure to any local configuration
containing two atoms separated by $k=0,1,...$ 2-molecules. 
Particles from neighboring
pairs are moved to each other and $k+1$ 2-molecules are created. 
Then, it appears that the internal energy of all considered particles
strictly decreases and the change is bounded from above
by the interaction energy between two atoms taken with the minus sign
while the interaction energy
of considered particles with all the rest does not increase. 
In this way we proved the following statement:

Consider a lattice gas (\ref{H}) with an interaction energy $V$
such that $V(1) \leq 0$. Among configurations that do not contain
$n$-molecules with $n \geq 3$, the lowest-energy configurations consist
exclusively of $2$-molecules.

The actual proofs of the two results, discussed above, are fairly technical
and for this reason they are included in a separate paper
\cite{jedmie}.

Combining the above two results we obtain that ground-state configurations
of our model contain only 2-molecules. It is easy to see that the effective
interaction between these molecules is strictly convex - it is 
a sum of strictly convex interactions between particles of the molecules.
The latter observation enables us to formulate a theorem that
constitutes our main result.

\noindent {\bf Theorem} ($2$-molecule most homogeneous ground-state
configurations)

\noindent Consider a lattice gas (\ref{H}) with a nonconvex interaction energy
$V$ and a particle density $\rho \leq 1/2$.
If $-V(2) <  V(1) \leq 0$ and $V(2) \geq (7W+V(1))/2$, then the ground-state
configurations are the $2$-molecule most homogeneous configurations
of particle density $\rho$.

\noindent {\bf Corollary} (Molecular devil's staircase)

\noindent In a lattice gas (\ref{H}) whose interaction energy $V$ satisfies
the conditions given in the theorem,
the particle density versus the chemical potential of particles,
$\rho (\mu)$, exibits the complete devil's-staircase structure.

To summarize, we have proven rigorously the existence 
of the molecular complete devil's staircase in one-dimensional 
lattice-gas models with certain nonconvex long-range interactions.

As explained at the very begining, our main result, concerned with
one-dimensional lattice gases, applies directly to
a class of two-dimensional lattice gases modelling the adsorption of
adatoms on crystal surfaces. Such models have been
investigated numerically by Ishimura and Yamamoto \cite{ishimura}. 
In particular, they posed the question about the possibility of the
presence of a devil's staircase in such models.
Due to the symmetry of the considered Hamiltonians (\ref{H}),
with respect to space rotations by $\pi /2$,
the two-dimensional ground-state adatom configurations
are given by the oblique arrangements
of one-dimensional ground-state configurations determined above.
Namely, the adatom configurations at lattice lines in the $x$-direction
can be chosen as the one-dimensional ground-state configurations,
with the condition that, going in the positive $y$-direction,
the configuration of the next lattice line is shifted by one lattice
constant in the positive $x$-direction.
Now, knowing the solution of the ground-state problem, it would be
very interesting to investigate the low-temperature
stability of such structures.

There is also a numerical evidence \cite{guj} that certain nonconvex
interactions appear, as effective two-body interactions, 
in the one-dimensional spinless Falicov-Kimball model
\cite{falkim,brandt}; it is known that
this quantum model can be transformed into a classical lattice-gas
model with fairly complicated longe-range and many-body interactions
\cite{kennlieb,grumacleb}.

Another class of systems, where our ground-state results are
immediately applicable, consists of three-dimensional layered systems,
like those studied by Fisher and collaborators \cite{fisher}.
As in the two-dimensional case, it is highly desirable to study the
stability of ground-state orderings against thermal fluctuations.

\noindent {\bf Acknowledgments.} We would like to thank 
the Polish Committee for Scientific Research, 
for a financial support under the grant KBN 2P03A01511.
One of the authors (J.\ J.) is grateful to
Ch.\ Gruber for discussions of the obtained results and kind
hospitality during his stay in the Institut de Physique Th\'{e}orique
of the Ecole Polytechnique F\'{e}d\'{e}rale de Lausanne.

\end{multicols}
\end{document}